\documentclass[twocolumn]{article}
\usepackage[english]{babel}
\usepackage[utf8]{inputenc}
\usepackage[dvips]{graphicx}
\usepackage[]{amsmath}
\usepackage{authblk}

\newcommand{\eq}{{\rm eq}}
\newcommand{\kb}{k_{\rm B}}

\newcommand{\ud}{\mathrm{d}}

\title{Differential Landauer's principle}
\author{Léo Granger}
\author{Holger Kantz}
\affil{Max-Planck-Institut für Physik komplexer
Systeme, Nöthnitzerstr. 38, D-01187, Dresden, Germany}

\begin{document}

\maketitle

\begin{abstract}
Landauer's principle states that the erasure of information
must be a dissipative process. In this paper, we carefully
analyze the
recording and erasure of information on a physical memory.
On the one hand, we show that in order to record some
information, the memory has to be driven out of equilibrium.
On the other hand, we derive a differential version of
Landauer's principle: We link the rate at which entropy is
produced at every time of the erasure process
to the rate at which information is erased.
\end{abstract}

\section{Introduction}

The fundamental role of information in thermodynamics is now
well established.
On the one hand, it is possible to convert information into
useful work and use it as a ``fuel'' for some heat engine 
\cite{szilard_uber_1929,sagawa_nonequilibrium_2012}.
On the other hand, information itself must be recorded on
some physical memory. Information processing is
implemented via physical transformations operating on the
memory and is thereby accompanied by thermodynamic costs
\cite{landauer_irreversibility_1961,leff_maxwells_2002,berut_experimental_2012,orlov_experimental_2012,granger_thermodynamic_2011,sagawa_minimal_2009,jacobs_quantum_2012,sagawa_fluctuation_2012-1,mandal_work_2012,horowitz_imitating_2012}.
The paradigmatic result concerning the costs of information
processing is Landauer's
erasure principle \cite{landauer_irreversibility_1961}.
It is a statement about the thermodynamics of
information erasure.
In its original formulation, it states that
the
resetting to zero of a random bit necessarily leads to the
dissipation of $\kb T \log 2$ of energy, where $\kb$ is
Boltzmann's constant and $T$ is the temperature of the
environment.
Landauer's principle was celebrated as the resolution
Maxwell's demon's paradox
\cite{bennett_thermodynamics_1982}.
However it is still a very controversial result,
see
\cite{norton_waiting_2011} and references
therein.
In this paper, we carefully analyze the processes of
recording and erasing information on a physical memory in a
simple but generic scenario and we
precisely derive the thermodynamic costs of these processes. 
While the problem usually addressed in the literature is the
operation ``reset to zero'' performed on isolated one-bit
memories
\cite{shizume_heat_1995,fahn_maxwells_1996,piechocinska_information_2000,plenio_physics_2001,dillenschneider_memory_2009,berut_experimental_2012,orlov_experimental_2012},
here we propose to analyze the recording and erasure of some
information coming from an external source.
We use the framework of stochastic thermodynamics
\cite{gaveau_general_1997,seifert_entropy_2005,esposito_second_2011,esposito_stochastic_2012,seifert_stochastic_2012-1}
to model the memory and to derive the thermodynamics of
these processes.

We consider the following scenario:
A source randomly emits a symbol $\alpha_k$ out of
$N$ possible symbols,
$\alpha_1,\dots,\alpha_N$.
We wish to record which symbol appeared on some physical
memory and then we wish to erase it.
The memory should
be able to be in at least $N$ states: One for each
of the possible outcomes. However, it is convenient to have
one more state serving as the ``standard'' state: It is the
state of the memory 
when it is empty, i.e.~when nothing is recorded.
Recording
$\alpha_k$ simply means to drive the memory from the standard state
to state number $k$.  Erasing the content of the memory
means to drive it back to the standard state by a process
that is independent of the symbol stored.

In the following, we show that recording the information
amounts to correlate
the memory to the symbol emitted by the
source.
During the erasure process, these correlations gradually
decrease to zero. We show that the rate of entropy produced
is bounded from below by the rate at which the correlations
decrease all along the erasure process.
This is our second main result,
eq.~(\ref{eq:main2}), which can be viewed as a differential
version of Landauer's principle.
Our first main result, eq.~(\ref{eq:main1}) is a statement
about a different kind of costs for the correlations:
Although the recording can in principle be performed
reversibly, the
memory can only store some information if it is out of
equilibrium.

\section{Recording information on a physical system}

In order to simplify the discussion,
we use an over-damped particle subject to a random force of
thermal origin as a memory.
However, any system obeying the laws
of stochastic thermodynamics would do.
The particle is also subjected to a
force
$F(x,\lambda) = -\partial_x V(x,\lambda)$ 
derived from a conservative potential $V(x,\lambda)$.
The potential is time-dependent through the external control
parameter $\lambda$ that can be varied in order to control
the state of the memory.
The position of the particle evolves according to the Langevin equation:
\begin{equation}
	\dot x = \mu F(x,\lambda) + \xi(t),
	\label{eq:langevin}
\end{equation}
where $\mu$ is the mobility of the particle and $\xi(t)$ is
a Gaussian white noise with $\langle \xi(t)\rangle = 0$ and
$\langle \xi(t)\xi(t')\rangle = 2D\delta(t-t')$, $D$ being
the diffusion constant. When the medium is in equilibrium
at temperature $T$,
the diffusion constant and the mobility are related by the
Einstein relation $D = \kb T \mu$.
The
probability density $\rho(x,t)$ to find the particle at
position $x$ at time $t$ evolves according to the
Fokker-Planck equation:
\begin{equation}
	\partial_t \rho(x,t) = - \partial_x \left(
	\mu F(x,\lambda) \rho(x,t) - D\partial_x \rho(x,t)
	\right),
	\label{eq:fp}
\end{equation}
Because of thermal fluctuations, the position of the
particle is random and we cannot control it. However, through
a suitable choice of the time dependence of the control
parameter $\lambda(t)$ (the protocol), we can control
the distribution $\rho(x,t)$ describing the position of the
particle.
Hence, we will use such distributions 
to encode the different symbols $\alpha_k$.

Let $\{\phi_k\}$, $0\leq k\leq N$ be $N+1$ distributions
over the position of the particle: $\phi_0(x)$
is the standard state and for $1\leq k\leq N$, $\phi_k(x)$
encodes the symbol $\alpha_k$.
For the information to
be unambiguously recorded, we need the 
states encoding different symbols
to be perfectly distinguishable. This means that the
corresponding
distributions should not overlap:
For each position $x$ there
should be only one $k$ such that $\phi_k(x)$ is non zero.
In fact, the information is stored in the position of the
particle. If for a given position $x$,
$\phi_k(x)>0$ and
$\phi_{k'}(x)>0$, then observing the particle at $x$ we cannot
decide whether $\alpha_k$ or $\alpha_{k'}$ is
stored.

\begin{figure}
	\includegraphics{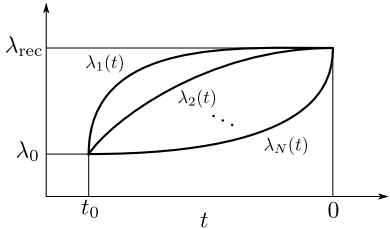}
	\caption{Recording protocol: From time $t = t_0 <0$
	to $t = 0$, the control parameter $\lambda$ is
	driven from $\lambda(t_0)=\lambda_0$ to $\lambda(0)
	= \lambda_{\rm rec}$ in a way that depends on $k$,
	such that the final state of the memory is $\phi_k$,
	the state encoding $\alpha_k$, the symbol that was
	emitted by the source.}
	\label{fig:recprot}
\end{figure}

The recording process is the following.
Assume that the symbol emitted by the source is $\alpha_k$.
Initially, the memory is in the
standard state $\rho(x,t_0) =\phi_0(x)$
and the control parameter takes stored
value $\lambda(t_0) = \lambda_0$.
From time $t=t_0 < 0$ to $t = 0$, the control
parameter is changed from its initial value $\lambda_0$ to
some final value $\lambda_{\rm rec}$ according to some
protocol
$\lambda_k(t)$.
The protocol 
depends on the symbol $\alpha_k$
that appeared and is such that the
final state of the memory is $\rho(x,0) =\phi_k(x)$.
However, once the information is recorded, we want to be
able to manipulate the memory without knowing the
information stored. Hence, the final value of the control
parameter, $\lambda_{\rm rec}$ should be the same for all
$k$ as illustrated on figure \ref{fig:recprot}.
As a  consequence, at the end of the recording process,  the
memory is out of equilibrium.
 In fact we want to allow 
it to be in one out of $N$ different states
for a single value of $\lambda$. But for each
value of $\lambda$, there is only one
equilibrium state given by the Boltzmann distribution:
\begin{equation}\label{eq:eq}
\rho_\eq(x,\lambda) = \exp\left(-\frac{V(x,\lambda) -
F_\eq(\lambda)}{\kb T}\right),
\end{equation}
where $\exp(-F_\eq(\lambda)/\kb T) = \int
\exp(-V(x,\lambda)/\kb T) \ud x$ is the partition function 
and $F_\eq(\lambda)$ the equilibrium free energy as a
function $\lambda$.

In stochastic thermodynamics, the distance of a non
equilibrium state $\rho$ to equilibrium $\rho_\eq$ is
quantified
by the Kullback-Leibler divergence, or relative entropy
between $\rho$ and $\rho_\eq$
\cite{vaikuntanathan_dissipation_2009,esposito_second_2011,deffner_information_2012}:
\begin{equation}
	D[\rho\|\rho_\eq] = \int \rho(x)
	\log\frac{\rho(x)}{\rho_\eq(x)}\ud x.
	\label{eq:kl}
\end{equation}
In fact, this quantity is
linked to the free energy of the non equilibrium state
$\rho$: $F[\rho] - F_\eq = \kb TD[\rho\|\rho_\eq] \geq 0$
\cite{gaveau_general_1997,esposito_second_2011,deffner_information_2012}. 
The relative entropy between two distributions is non
negative and it vanishes if and only if the two
distributions are identical \cite{cover_elements_2006}.

The first result
of this paper is the following inequality:
\begin{equation}
	\sum_k P_k D[\phi_k\|\rho_\eq] \geq H,
	\label{eq:main1}
\end{equation}
where  $\rho_{\rm eq}$ is the
final equilibrium distribution,
$P_k$ is the probability that the symbol $\alpha_k$
was emitted and
$H = -\sum_k P_k\log P_k$ is the Shannon entropy of
the source.
The latter quantifies our a priori uncertainty about the
symbol emitted by the source or the average information
provided by the observation of the symbol emitted
\cite{cover_elements_2006}.
Inequality
(\ref{eq:main1}) above means, that the expected distance
to equilibrium at the end of the recording process is
greater than the average information provided by the
emission of the symbol to be recorded. One can interpret $H$
as the amount of information stored in the memory.
Inequality (\ref{eq:main1}) can then be interpreted in the
following way: In order to record some information, the
system serving as a memory has to be driven out of
equilibrium and the distance to equilibrium has to be
greater than the information stored.
Nevertheless, the recording process can in principle be
performed reversibly in that there is no finite lower bound
to the entropy that has to be produced.

Inequality (\ref{eq:main1}) is very
general and does not depend on the structure of the memory.
The only assumptions are: (i) The macroscopic states of the
memory are given by distributions over the micro-states,
(ii) states of the memory encoding different symbols
should be perfectly distinguishable, (iii) there is a unique
equilibrium distribution.
Sometimes, due to technical restrictions, it might not be
possible to prepare the memory in non overlapping states and
hence it would not be possible to fulfill condition (ii).
In this case the information cannot
be completely unambiguously recorded, but
nevertheless, as we will see later, one can still
quantify the maximum amount of
information, $I_{\rm max} < H$, that can be stored and
inequality
(\ref{eq:main1}) will still be valid replacing $H$ by
$I_{\rm max}$ in the right hand
side. We will address this point later, after having
discussed the erasure process.

\section{Erasure: differential Landauer's principle}

The erasure is a process bringing the memory back to
its standard state
without making use of the information stored.
This is expressed by the fact that the erasure protocol does
not depend on $k$.
The initial value of the control parameter is $\lambda_{\rm
rec}$ and the initial state of the memory is the state
encoding the symbol that was emitted, i.e.~it is $\phi_k(x)$
with probability $P_k$. From time $t=0$ to time $t=t_1$ the
control parameter is driven from $\lambda_{\rm rec}$ back to
$\lambda_0$ in such a way that the final state is the
standard state $\phi_0(x)$.

Assume that we stop the process at some intermediate time
$t$.
We would like to
address the following issues: What is the information erased
until that time? Or alternatively,
what is the information still contained in
the memory? And what is the minimum amount of entropy produced
until that time?

In order to illustrate the information loss,
consider the following situation:
At time $t$, we see the particle at position $x$ and from this
knowledge, we would like to infer which symbol was stored.
The probability $P(k|x;t)$, that the symbol $\alpha_k$ was
stored is
given by Bayes rule:
\begin{equation}
	P(k|x;t) = \frac{\rho_k(x,t) P_k}{\rho_{\rm m}(x,t)},
	\label{eq:bayes}
\end{equation}
where $\rho_k(x,t)$ is
the distribution of the particle's position at time $t$ 
if $\alpha_k$ was stored and
$\rho_{\rm m}(x,t) = \sum_k P_k \rho_k(x,t)$ is the
marginal distribution of the position of the particle at
time $t$. Its presence in the expression above ensures that
the probability distribution $P(k|x;t)$ is normalized.
Concretely, $\rho_k(x,t)$ is obtained by propagating
$\phi_k(x)$ with the Fokker-Planck equation with the erasure
protocol $\lambda(t)$.
Initially, at $t= 0$, $P(k|x;0) = 0$ or
1 depending on whether $x$ belongs to the support of
$\phi_k$ of not: The position of the particle contains the
complete information about the symbol that was emitted.
At intermediate time $t$, the
$\rho_k(x,t)$ might overlap and we will have some
uncertainty about the symbol that was stored upon seeing
the particle at position $x$.
This uncertainty  is quantified by the Shannon entropy of
the probability distribution $P(k|x;t)$:
\begin{equation}
	h_{\rm er}(x,t) = - \sum_k
	P(k|x;t) \log P(k|x;t).
	\label{eq:condentkx}
\end{equation}
The total information erased until time $t$ is the average
uncertainty about the symbol that was stored upon knowing
the position of the particle at time $t$:
\begin{equation}
	H_{\rm er}(t) = \int
	\rho_{\rm m}(x,t)h_{\rm er}(x,t)\ud x.
	\label{eq:condentk}
\end{equation}
At the beginning of the erasure process, no information is
yet erased and 
$H_{\rm er}(0) = 0$. At the end of the process,
knowing the position of the particle does not reduce our
uncertainty about the symbol that had been emitted and
$H_{\rm er}(t_1) = H$. The information $I(t)$ still contained in
the memory is the
reduction in uncertainty about the symbol that was stored
upon knowing the position of the particle at time $t$:
\begin{equation}
	I(t)
= H - H_{\rm er}(t).
	\label{eq:info}
\end{equation}
This is just the {\em mutual information} between the
position of the particle and and the symbol originally
recorded. It is a measure of how much information 
the position of the
particle at time $t$ can still provide about the symbol 
originally stored
\cite{cover_elements_2006}. During the erasure process, it decreases from
$H$ to 0 as the information is gradually erased.

The second main result of this paper is the following:
The rate of entropy production is bounded from below by the
rate of information erasure:
\begin{equation}
	\dot S^{\rm irr} \geq \kb \dot H_{\rm er}= -\kb \dot I.
	\label{eq:main2}
\end{equation}
We call this result
the differential Landauer's principle.
In fact, it is a precise and general statement of the
thermodynamic costs of information erasure.
Integrating
the
equation above
between times $t$ and $t'$, with $0 \leq t \leq
t'\leq t_1$, we get a lower bound for the irreversible
entropy production between time $t$ and $t'$:
\begin{equation}
	\Delta S^{\rm irr}(t,t') \geq \kb\left(  I(t) -
	I(t')
	\right).
	\label{eq:landint}
\end{equation}
Hence, the minimum amount of entropy produced is directly
linked to the loss of correlation between the position of
the particle and the symbol initially stored.
Setting $t = 0$ and $t' = t_1$ in the equation
(\ref{eq:landint}) above
gives a general integral version of Landauer's principle:
\begin{equation}
	\Delta S^{\rm irr} \geq \kb H,
	\label{eq:tradland}
\end{equation}
linking the total amount $\Delta S^{\rm irr}$ of entropy  
produced to the amount $H$
of information erased.

In general, we might not be able to prepare the memory in
non overlapping states $\{\phi_k\}$. In this case, the
information cannot be fully reliably stored. However, we can
still quantify the maximum amount of information that we can
store.
It is given by eq.~(\ref{eq:info}) for $t=0$:
$I_{\rm max} =
I(0) = H - H_{\rm er}(0) < H$, where $H_{\rm er}(0) > 0$ due
to the overlapping.
In this case, inequality (\ref{eq:main1}) still holds
replacing $H$ by $I_{\rm max}$ in the
right hand side. Moreover,
the lower bound to the total entropy produced during the
erasure process, obtained by integrating inequality
(\ref{eq:main2}), is reduced: $\Delta S^{\rm irr} \geq \kb
I_{\rm max}$, implying that we do not have to pay for the
information we were not able to record.

\section{Proof of equations (\ref{eq:main1}) and
(\ref{eq:main2})}

We now prove our two main results, inequalities
(\ref{eq:main1}) and (\ref{eq:main2}).
The information contained in the memory at time $t$ of the
erasure process can be expressed as:
\begin{equation}
	I(t) = \sum_k P_k D[\rho_k(t)\|\rho_{\rm m}(t)].
	\label{eq:infokl}
\end{equation}
Using this expression, it is easy to show that the average
distance to equilibrium satisfies:
\begin{equation}
	\sum_k P_k D[\rho_k(t)\|\rho_\eq(\lambda(t))] = I(t)
	+ D[\rho_{\rm m}(t)\|\rho_\eq(\lambda(t))].
	\label{eq:proofmain1}
\end{equation}
At time $t=0$, this expression implies inequality
(\ref{eq:main1}), since $I(0) = H$, and the second term in
the right hand side is non negative at all times. Actually,
eq.~(\ref{eq:proofmain1}) is a generalization of
inequality (\ref{eq:main1}). It relates the average distance
to equilibrium to the information still contained in the
memory at any time of the erasure process.

At time $t$ of the erasure process, if $\alpha_k$ was
stored, the state of the memory
is $\rho_k(x,t)$ introduced earlier. The thermodynamic
entropy of the
memory is given by the Shannon-Gibbs formula:
\begin{equation}
	S_k(t) = -\kb \int \rho_k(x,t)\log \rho_k(x,t)\ud x.
	\label{eq:shannongibbs}
\end{equation}
Its variation satisfies the Clausius relation
\cite{esposito_second_2011,van_den_broeck_three_2010}:
\begin{equation}
	\dot S_k(t) = \frac{\dot Q_k(t)}{T} +
	\dot S_k^{\rm irr}(t),
	\label{eq:clausiusk}
\end{equation}
where 
$\dot Q_k(t) = \int V(x,\lambda(t))\partial_t
\rho_k(x,t)\ud x$ is the heat flux to the particle at time
$t$ and $\dot S_k^{\rm irr}(t) \geq 0$ is the rate at which
entropy is irreversibly produced.

Using eq.~(\ref{eq:infokl}), we get the
following relation
between the information $I(t)$ and the expected entropy $S(t)
= \sum_k P_k S_k(t)$ of the memory:
\begin{equation}
	S(t) = S_{\rm m}(t) - \kb I(t),
	\label{eq:entinfo}
\end{equation}
where $S_{\rm m}(t) = -\kb \int \rho_{\rm m}(x,t) \log
\rho_{\rm m}(x,t)\ud x$ is the entropy of the marginal
distribution $\rho_{\rm m}(x,t)$. Since the Fokker-Planck
eq.~(\ref{eq:fp}) is linear and since the marginal
distribution is a linear combination of solutions of this
equation, it satisfies this equation as well. As a
consequence, the
variations of $S_{\rm m}(t)$ also satisfy some Clausius
relation similar to eq.~(\ref{eq:clausiusk}):
\begin{equation}
	\dot S_{\rm m}(t) = \frac{\dot Q_{\rm m}(t)}{T} +
	S_{\rm m}^{\rm irr}(t),
	\label{eq:clausiusm}
\end{equation}
where $\dot Q_{\rm m}(t) = \int V(x,\lambda(t)) \partial_t
\rho_{\rm m}(x,t)\ud x$ and $S^{\rm irr}_{\rm m}(t) \geq 0$.

Noting that $\dot Q_{\rm m}(t) = \sum_k P_k \dot Q_k(t)$ and
inserting equations (\ref{eq:clausiusk}) and
(\ref{eq:clausiusm}) into eq.~(\ref{eq:entinfo}) yields
for the average entropy production rate $\dot S^{\rm irr}(t)
= \sum_k P_k \dot S_k^{\rm irr}(t)$:
\begin{equation}
	\dot S^{\rm irr}(t) = 
	\dot S_{\rm m}^{\rm irr}(t) - \kb \dot
	I(t).
	\label{eq:proofmain2}
\end{equation}
This proves eq.~(\ref{eq:main2}) since $\dot S_{\rm
m}^{\rm irr}(t) \geq 0$.

We used an over-damped Brownian particle as a memory in
order to present our results clearly. However, one could use
other types of systems obeying the laws of stochastic
thermodynamics as a memory. In particular, systems with
discrete phase space evolving according to master equations
such as the systems considered in
\cite{esposito_three_2010-1,esposito_stochastic_2012} 
yield the same results, where it is essential that they
fulfill the Clausius relation, eq.~(\ref{eq:clausiusk}).
In fact in the following we will present the recording and
erasure of the outcome of a binary random variable on a
memory with two discrete micro-states.

\section{Example: a two states system as a memory}

As an example, we consider the recording and erasure of the
result of a binary random variable. Let $Y$ be a random
variable with two possible outcomes. In analogy with the
tossing of a biased coin, let ``head'' and ``tail'' be the
two possible outcomes
appearing respectively with probability $P$ and $1-P$.
We want to record the outcome of $Y$ and then
erase it. The amount of information we want to store
is given by the Shannon entropy of $Y$, $H(P) = -P\log P -
(1-P)\log(1-P)$.

\begin{figure}
	\includegraphics{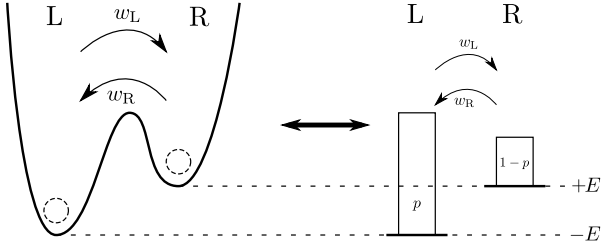}
	\caption{A very simple model for a memory: A system
	that can make thermally activated transition between
	two states ``L'' and ``R''. This system can be
	controlled through the energy difference between the
	two states. Its macroscopic state is given by the
	probability $p$ to occupy the state ``L''. This
	system could represent a Brownian particle in a
	double well potential, that can jump from one well
	to the other because of thermal fluctuations.}
	\label{fig:2states}
\end{figure}

The system serving as a memory is a two states system in
contact with an equilibrium heat bath at temperature $T$.
This system could represent an over-damped Brownian particle in a
double well potential as depicted in figure
\ref{fig:2states}.
\begin{figure}
	\includegraphics{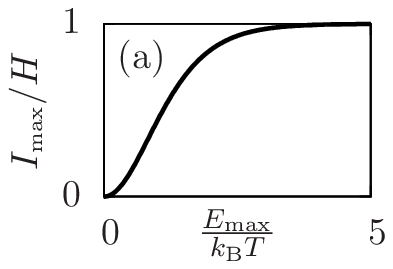}
	\hspace{0.3cm}
	\includegraphics{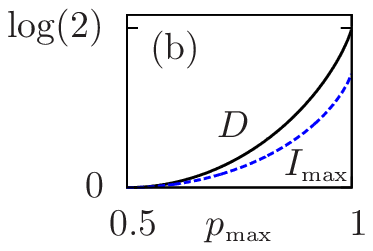}
	\caption{(a) $I_{\rm max}/H$ as a function of
	$E_{\rm max}/\kb T$ for $P=0.8$. The result is very
	similar for other values of $P$.
	(b) Kullback-Leibler divergence
	$D = PD(p_{\rm h}\|p_\eq) + (1-P)D(p_{\rm t}\|p_\eq)$
	to equilibrium at the
	end of the recording process and maximum information
	$I_{\rm max}$
	stored in the memory as a function of $p_{\rm max}$.
	}
	\label{fig:imax}
\end{figure}
In the following, we will again speak
of a ``particle'' that can ``jump'' between two ``wells''
keeping in mind that the memory could be any two states
system in contact with a heat bath that can make transitions
between the two states.
Let ``L'' and ``R'' label the two wells
and let $E_{\rm L} = - E$ and $E_{\rm R} = + E$ be their
respective energies. We assume that we are able to control the
energy difference $\Delta E = E_{\rm R}-E_{\rm L} = 2E$
between the two wells by applying a force field.
Hence, our control parameter is $E$.

At any time the particle might jump from
one well to the other due to thermal fluctuations of the
heat bath. The probability per unit time, that the particle
jumps from the right to the left well (from the left to the
right well) is given by
Kramer's rate $w_{\rm R} = \tau^{-1}\exp\left(
E/\kb T
\right)$ ($w_{\rm L} = \tau^{-1}\exp\left(
-E/\kb T
\right)$), where $\tau$ is a time linked to the height of the
potential barrier between the two wells.
The macroscopic state of the memory is fully described by
the probability $p$ that the particle is in the left well.
The latter evolves in time according to the following master
equation:
\begin{equation}
	\dot p = -w_{\rm L} p + w_{\rm R}\left( 1-p
	\right).
	\label{eq:master}
\end{equation}
The rates satisfy detailed
balance $w_{\rm R}/w_{\rm L} = \exp(-\Delta E
/\kb T)$ and
the equilibrium is given by the Boltzmann
factor
\begin{equation}
	p_{\rm eq}(E) = \exp\left( -\frac{E_{\rm L} - F}{ \kb
	T} \right),
	\label{eq:eq2states}
\end{equation}
where the equilibrium free energy $F(E)$ is linked to the partition
function $Z(E) = \exp(-F(E)/\kb T) = 2\cosh( E/\kb T)$.
The thermodynamic entropy of the memory is given by $S(p) =
-\kb ( p\log p + (1-p) \log (1-p))$ and 
the
heat received per unit time is given by $\dot Q = \dot
p \Delta E = 2\dot pE$. The
instantaneous entropy production rate is given by the
Clausius relation $\dot S^{\rm irr} = \dot S - \dot
Q / T\geq 0$.

Now, we need to decide which states we use to
encode the different possible outcomes of $Y$. The idea is
to say that the left well encodes one of the outcomes, say
``head'' and the right well the other. In other words, we
want the state
$p_{\rm h} = 1$ to encode ``head'' and the state
$p_{\rm t} = 0$ to encode ``tail''. However, this supposes
that we are able to create an infinite energy difference
between the two wells.
In the following,
we assume that there is a maximum value
$E_{\rm max}\geq 0$, such that $|E| \leq E_{\rm max}$. Then
the best we can do is drive the memory to $p_{\rm max} =
p_\eq(E_{\rm max})$ or $1-p_{\rm max} = p_\eq(-E_{\rm
max})$. Hence, we will use $p_{\rm h} = p_{\rm max}$ to
encode ``head'' and $p_{\rm t} = 1-p_{\rm max}$ to encode
``tail''. Introducing $p_{\rm m} = P p_{\rm h} + (1-P)
p_{\rm t}$, the marginal probability for the particle
to occupy the left well at the end of the recording process,
we can compute the maximum amount of information that can be
stored in this memory using eq.~(\ref{eq:infokl}):
\begin{equation}
	I_{\rm max} = P D(p_{\rm h}\|p_{\rm m}) +
	(1-P)D(p_{\rm t}\|p_{\rm m}),
	\label{eq:imax2states}
\end{equation}
where $D(p\|q) =
p\log\frac{p}{q} + (1-p)\log\frac{1-p}{1-q}$ is the 
Kullback-Leibler divergence
between two distributions over a binary random
variable.
Figure \ref{fig:imax} (a) shows $I_{\rm max}/H$ as a function of
$E_{\rm max}/\kb T$.  For $E_{\rm max}$ of the order of $\kb T$,
the maximum information content is clearly smaller than $H$.
For $E_{\rm max} \gg \kb T$, $I_{\rm max} \simeq H$.

Now that we have the states encoding the different possible
outcomes, we have to specify the standard state.
Traditionally in the literature, the standard state is the
state ``the particle is in the left (or right) well''
\cite{landauer_irreversibility_1961,dillenschneider_memory_2009,berut_experimental_2012}.
However, we can use any distribution over the two
wells as
the standard state. A simple choice is the equidistribution:
$p_0 = 1/2$.

\begin{figure}
	\includegraphics[scale=1.1]{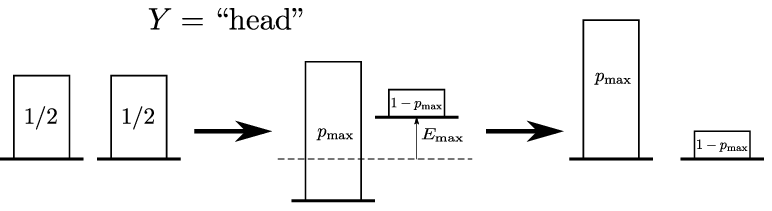}
	\includegraphics[scale=1.1]{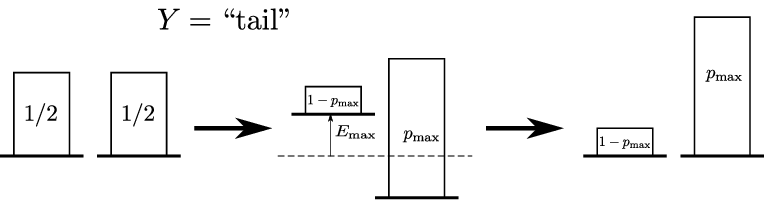}
	\caption{Schematic picture of the recording process
	in the two possible scenarios. The control parameter
	$E$ is driven to $E_{\rm max}$ or $-E_{\rm max}$
	depending on the outcome of $Y$. Once equilibrium is
	reached, $E$ is quickly driven back to 0.}
	\label{fig:recording2states}
\end{figure}

The
recording process is schematically sketched on figure
\ref{fig:recording2states}.
If ``head'' appeared,
the parameter $E$ is driven from $0$ to $E_{\rm max}$ and the
memory is let to relax towards equilibrium $p_{\rm h} = p_{\rm max}$ and if
``tail'' appeared we drive $E$ to $-E_{\rm max}$ and let
the memory relax towards $p_{\rm t} = 1-p_{\rm max}$. At the end of
the recording process, we instantaneously drive $E$ back to zero,
so that it has the same value in the two cases.
As can be seen on figure \ref{fig:imax} (b), at the end of
the recording process, the Kullback-Leibler
divergence  to the equilibrium state is greater than the
maximum information stored.

The erasure process is
very simple: We keep $E = 0$ and 
simply let the memory relax towards equilibrium.
\begin{figure}
	\includegraphics{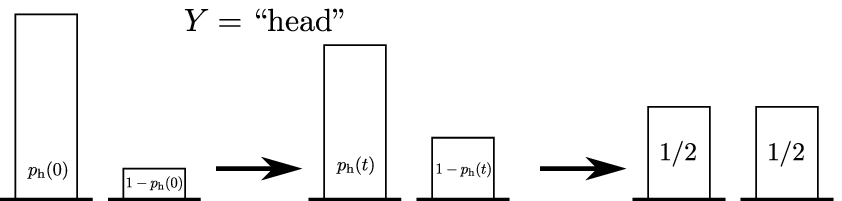}
	\vspace{0.3cm}

	\includegraphics{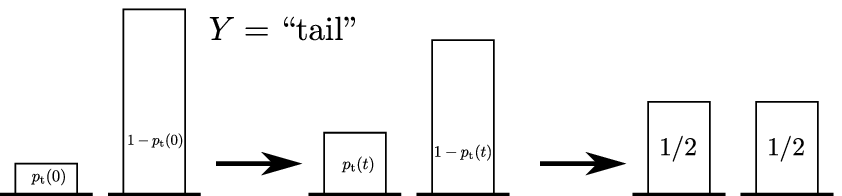}

	\vspace{0.5cm}
	\begin{tabular}{p{0.16\textwidth}p{0.16\textwidth}p{0.10\textwidth}}
		$ I = I_{\rm max}$ & $ I < I_{\rm max} $ & $
		I = 0$
	\end{tabular}
	\caption{Schematic picture of the erasure process.
	The memory, initially out of equilibrium is simply
	let to equilibrate. The information is erased as the
	overlap between $p_{\rm h}(t)$ and $p_{\rm t}(t)$
	increase.}
	\label{fig:erasing2states}
\end{figure}
\begin{figure}
	\includegraphics{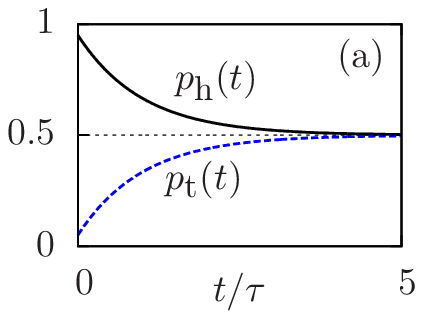}
	\hspace{0.2cm}
	\includegraphics{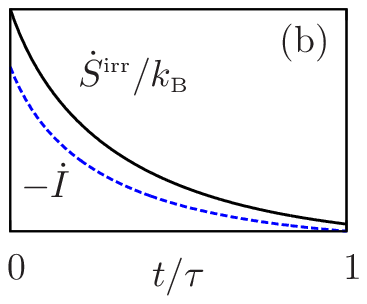}

	\vspace{0.3cm}
	\includegraphics{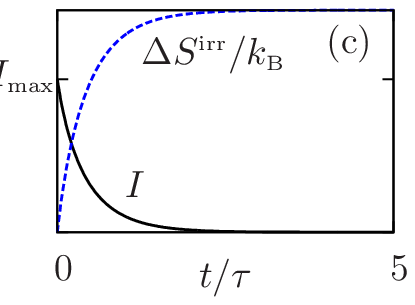}
	\includegraphics{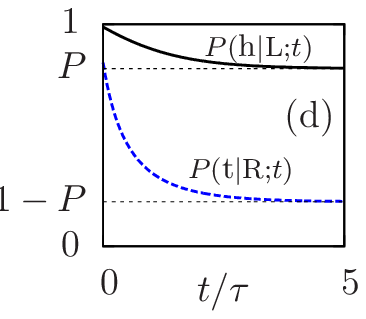}
	\caption{Time evolution of various quantities during
	the erasure process for $p_{\rm max} = 0.95$ and $P
	= 0.8$:(a) Evolution of the state of the memory
	during the erasure process. (c) Information
	content and total entropy produced during the
	erasure process. (d) Conditional probability that
	``head'' (``tail'') was recorded observing the
	particle in the left (right) well at time $t$ of the
	erasure process.}
	\label{fig:traj}
\end{figure}
The erasure process is schematically sketched on figure
\ref{fig:erasing2states}.
If ``head'' was recorded, the initial state of the
memory is $p_{\rm h}(0) = p_{\rm max}$ and if ``tail'' 
was recorded, the initial state of the memory is $p_{\rm t}(0) =
1-p_{\rm max}$. The initial information contained in the
memory is $I(0) = I_{\rm max}$. As time goes on,
$p_{\rm h}(t)$ and $p_{\rm t}(t)$ converge towards
equilibrium, which is also the standard state
$p_\eq = p_0 = 1/2$. During this process, the
information decreases and reaches 0{} in the limit $t
\gg\tau$, where $p_{\rm h}(t) = p_{\rm t}(t) = p_0$, see
figure \ref{fig:traj}. The rate of entropy production
$\dot S^{\rm irr}/\kb$ (in units of $\kb$) is
indeed greater than the rate of information erasure $-\dot
I$ at all times.
Figure \ref{fig:traj} (d) shows the time evolution of the
conditional probabilities
$P({\rm h}|{\rm L};t) = P p_{\rm h}(t)/p_{\rm m}(t)$
that ``head'' was stored observing the particle in the left well
 and  $P({\rm t}|{\rm R};t) = (1-P)(1- p_{\rm
t}(t))/(1-p_{\rm m}(t))$ that
``tail'' was stored observing the particle in the right well
at time $t$. At the beginning of the erasure process, they
are close to 1. During the erasure process, they decrease
respectively towards $P$ and $1-P$, the a priori
probabilities that ``head'' and ``tail'' were stored
respectively. As $P({\rm h}|{\rm L};t)$ and $P({\rm t}|{\rm
R};t)$ decrease, it becomes more and more difficult to know
whether ``head'' or ``tail'' was stored upon knowing in
which well the particle is: The position of the particle
gradually looses the information about the symbol that was
initially stored.

\section{Conclusion}

We have presented the thermodynamics of recording and erasing
information on a
physical memory within the framework of stochastic
thermodynamics.
Recording some information means to correlate
the memory to the external source and
the information contained in the memory is quantified
by the mutual information between the symbol recorded
and the micro-state of the memory.
On the one hand, in order to store some information, the
memory must be out of equilibrium and its average
distance to equilibrium (measured in terms of
Kullback-Leibler divergence) must be greater than the
information stored.
On the other hand, when the information is erased, entropy
is produced at a rate greater than the information erasure
rate (up to a factor $\kb$).
This result is a differential generalization of Landauer's
principle, precisely stating the thermodynamic costs of
erasing information without making reference to
manipulations on one bit
memories.

\bibliographystyle{unsrt}
\bibliography{biblio}{}

\begin{thebibliography}{10}

\bibitem{szilard_uber_1929}
L.~Szilard.
\newblock über die entropieverminderung in einem thermodynamischen system bei
  eingriffen intelligenter wesen.
\newblock {\em Zeitschrift für Physik}, 53(11):840--856,
  1929.

\bibitem{sagawa_nonequilibrium_2012}
Takahiro Sagawa and Masahito Ueda.
\newblock Nonequilibrium thermodynamics of feedback control.
\newblock {\em Physical Review E}, 85(2):021104, February 2012.

\bibitem{landauer_irreversibility_1961}
R.~Landauer.
\newblock Irreversibility and heat generation in the computing process.
\newblock {\em {IBM} Journal of Research and Development}, 5(3):183 --191, July
  1961.

\bibitem{leff_maxwells_2002}
Harvey~S Leff and Andrew~F Rex.
\newblock {\em Maxwell's demon 2 : entropy, classical and quantum information,
  computing}.
\newblock Institute of Physics, Bristol, 2002.

\bibitem{berut_experimental_2012}
Antoine Bérut, Artak Arakelyan, Artyom Petrosyan, Sergio Ciliberto, Raoul
  Dillenschneider, and Eric Lutz.
\newblock Experimental verification of landauer's principle linking
  information and thermodynamics.
\newblock {\em Nature}, 483(7388):187--189, March 2012.

\bibitem{orlov_experimental_2012}
Alexei~O. Orlov, Craig~S. Lent, Cameron~C. Thorpe, Graham~P. Boechler, and
  Gregory~L. Snider.
\newblock Experimental test of landauer's principle at the
  sub-$k_{\text B} T$ level.
\newblock {\em Japanese Journal of Applied Physics}, 51:06FE10, 2012.

\bibitem{granger_thermodynamic_2011}
Léo Granger and Holger Kantz.
\newblock Thermodynamic cost of measurements.
\newblock {\em Physical Review E}, 84(6):061110, December 2011.

\bibitem{sagawa_minimal_2009}
Takahiro Sagawa and Masahito Ueda.
\newblock Minimal energy cost for thermodynamic information processing:
  Measurement and information erasure.
\newblock {\em Physical Review Letters}, 102(25):250602, June 2009.

\bibitem{jacobs_quantum_2012}
Kurt Jacobs.
\newblock Quantum measurement and the first law of thermodynamics: The energy
  cost of measurement is the work value of the acquired information.
\newblock {\em Physical Review E}, 86(4):040106, October 2012.

\bibitem{sagawa_fluctuation_2012-1}
Takahiro Sagawa and Masahito Ueda.
\newblock Fluctuation theorem with information exchange: Role of correlations
  in stochastic thermodynamics.
\newblock {\em Physical Review Letters}, 109(18):180602, November 2012.

\bibitem{mandal_work_2012}
Dibyendu Mandal and Christopher Jarzynski.
\newblock Work and information processing in a solvable model of maxwell’s
  demon.
\newblock {\em Proceedings of the National Academy of Sciences},
  109(29):11641--11645, July 2012.

\bibitem{horowitz_imitating_2012}
Jordan~M. Horowitz, Takahiro Sagawa, and Juan M.~R. Parrondo.
\newblock Imitating chemical motors with optimal information motors.
\newblock {\em {arXiv:1210.6448}}, October 2012.

\bibitem{bennett_thermodynamics_1982}
Charles~H. Bennett.
\newblock The thermodynamics of computation—a review.
\newblock {\em International Journal of Theoretical Physics}, 21(12):905--940,
  1982.

\bibitem{norton_waiting_2011}
John~D. Norton.
\newblock Waiting for landauer.
\newblock {\em Studies in History and Philosophy of Science Part B: Studies in
  History and Philosophy of Modern Physics}, 42(3):184--198, August 2011.

\bibitem{shizume_heat_1995}
Kousuke Shizume.
\newblock Heat generation required by information erasure.
\newblock {\em Physical Review E}, 52(4):3495--3499, October 1995.

\bibitem{fahn_maxwells_1996}
Paul~N. Fahn.
\newblock Maxwell's demon and the entropy cost of information.
\newblock {\em Foundations of Physics}, 26(1):71--93, January 1996.

\bibitem{piechocinska_information_2000}
Barbara Piechocinska.
\newblock Information erasure.
\newblock {\em Physical Review A}, 61(6):062314, May 2000.

\bibitem{plenio_physics_2001}
M.~B. Plenio and V.~Vitelli.
\newblock The physics of forgetting: Landauer's erasure principle and
  information theory.
\newblock {\em Contemporary Physics}, 42(1):25--60, 2001.

\bibitem{dillenschneider_memory_2009}
Raoul Dillenschneider and Eric Lutz.
\newblock Memory erasure in small systems.
\newblock {\em Physical Review Letters}, 102(21):210601, May 2009.

\bibitem{gaveau_general_1997}
Bernard Gaveau and {L.S.} Schulman.
\newblock A general framework for non-equilibrium phenomena: the master
  equation and its formal consequences.
\newblock {\em Physics Letters A}, 229(6):347--353, June 1997.

\bibitem{seifert_entropy_2005}
Udo Seifert.
\newblock Entropy production along a stochastic trajectory and an integral
  fluctuation theorem.
\newblock {\em Physical Review Letters}, 95(4):040602, July 2005.

\bibitem{esposito_second_2011}
M.~Esposito and C.~Van~den Broeck.
\newblock Second law and landauer principle far from equilibrium.
\newblock {\em {EPL} (Europhysics Letters)}, 95(4):40004, August 2011.

\bibitem{esposito_stochastic_2012}
Massimiliano Esposito.
\newblock Stochastic thermodynamics under coarse graining.
\newblock {\em Physical Review E}, 85(4):041125, April 2012.

\bibitem{seifert_stochastic_2012-1}
Udo Seifert.
\newblock Stochastic thermodynamics, fluctuation theorems and molecular
  machines.
\newblock {\em Reports on Progress in Physics}, 75(12):126001, December 2012.

\bibitem{vaikuntanathan_dissipation_2009}
S.~Vaikuntanathan and C.~Jarzynski.
\newblock Dissipation and lag in irreversible processes.
\newblock {\em {EPL} (Europhysics Letters)}, 87(6):60005, September 2009.

\bibitem{deffner_information_2012}
Sebastian Deffner and Eric Lutz.
\newblock Information free energy for nonequilibrium states.
\newblock {\em {arXiv:1201.3888}}, January 2012.

\bibitem{cover_elements_2006}
Thomas~M. Cover and Joy~A. Thomas.
\newblock {\em Elements of Information Theory}.
\newblock John Wiley \& Sons, July 2006.

\bibitem{van_den_broeck_three_2010}
Christian Van~den Broeck and Massimiliano Esposito.
\newblock Three faces of the second law. {II.} fokker-planck formulation.
\newblock {\em Physical Review E}, 82(1):011144, July 2010.

\bibitem{esposito_three_2010-1}
Massimiliano Esposito and Christian Van~den Broeck.
\newblock Three faces of the second law. I. master equation formulation.
\newblock {\em Physical Review E}, 82(1):011143, July 2010.

\end{thebibliography}

\end{document}